\documentclass[a4paper]{article}

\usepackage{cite}             
\usepackage{graphicx}
\usepackage[margin=1.5in]{geometry}

\begin{document}


\title{The KM3NeT Neutrino Telescope and the potential of a neutrino beam from Russia to the Mediterranean Sea}

\author{Dmitry Zaborov\\{\it CPPM, Aix-Marseille Universit\'e, CNRS/IN2P3, 13288 Marseille, France} \and for the KM3NeT Collaboration}

\date{}
\maketitle


\begin{abstract}
KM3NeT is a new generation neutrino telescope currently under construction at two sites in the Mediterranean Sea. At the Capo Passero site, 100 km off-shore Sicily, Italy, a volume of more than one cubic kilometre of water will be instrumented with optical sensors. This instrument, called ARCA, is optimized for observing cosmic sources of TeV and PeV neutrinos. The other site, 40 km off-shore Toulon, France, will host a much denser array of optical sensors, ORCA. With an energy threshold of a few GeV, ORCA will be capable to determine the neutrino mass hierarchy through precision measurements of atmospheric neutrino oscillations. In this contribution, we review the scientific goals of KM3NeT and the status of its construction. We also discuss the scientific potential of a neutrino beam from Protvino, Russia to ORCA. We show that such an experiment would allow for a measurement of the CP-violating phase in the neutrino mixing matrix. To achieve a sensitivity competitive with that of the other planned long-baseline neutrino experiments such as DUNE and T2HK, an upgrade of the Protvino accelerator complex will be necessary.
\end{abstract}

\section{Introduction}
The KM3NeT research infrastructure \cite{km3net_loi} includes two large underwater neutrino detectors in the Mediterranean Sea: ORCA (Oscillation Research with Cosmics in the Abyss) and ARCA (Astroparticle Research with Cosmics in the Abyss). Both detectors use the same operation principle and instrumentation technique, each optimally configured for a specific energy range. ORCA is optimized for the study of atmospheric neutrino oscillations in the energy range between 3 GeV and 20 GeV, with the primary goal to determine the neutrino mass hierarchy. When completed, ORCA will occupy a volume of 8~Mt of water. ARCA is optimized for the purposes of neutrino astronomy in the TeV--PeV energy range and will occupy a volume of more than 1~km$^3$ (1~Gt). Both ORCA and ARCA are now under construction. The ORCA site is located 40~km off-shore Toulon, France, 2450~m below the sea level. The ARCA site is 100~km off-shore Capo Passero, Sicily, Italy, 3500~m below the sea level.

This document is organised as follows. Section 2 introduces the KM3NeT detector technology. Sections 3 and 4 introduce the science programmes of ARCA and ORCA, respectively. Section 5 discusses the scientific case for an accelerator neutrino beam from Protvino, Russia, to ORCA. Conclusions are presented in Section 6.

\section{KM3NeT technology, operation principle, and current status}

ORCA and ARCA detect neutrinos through the detection of the Cherenkov light induced by secondary particles emerging from interactions of the neutrinos with the sea water.
Both ORCA and ARCA consist of several thousand Digital Optical Modules (DOMs). Each DOM contains 31 photomultiplier tubes, which detect the Cherenkov light, and associated electronics, all housed in a pressure-resistant glass sphere. The DOMs are arranged in vertical structures, called Detection Units (DU), with 18 DOMs on each DU. The DUs are anchored to the seabed and connected to the shore station via an underwater cable network. In the case of ORCA, the vertical spacing between the DOMs along the DU is 9~m, and the DUs are installed on the seabed with an average spacing of 23~m. When completed, ORCA will comprise 115 DUs. For ARCA, the vertical spacing between the DOMs is 36~m and the horizontal spacing between the DUs is 90~m. Upon completion of the second construction phase, ARCA will comprise 230 DUs grouped in two building blocks of 115 DUs each.

\section{Neutrino astronomy with ARCA}

The main scientific objective of KM3NeT-ARCA is the detection of high-energy neutrinos of cosmic origin, in particular from sources in our Galaxy \cite{km3net_loi}. The preferred search strategy is to identify upward-moving muons, which unambiguously indicates neutrino reactions, since only neutrinos can traverse the Earth without being absorbed. The expected angular resolution is better than 0.3$^{\circ}$ at E $>$ 10 TeV. It should be noted that most of the potential Galactic sources are in the Southern sky. In this regard, ARCA is complementary to IceCube which is located at the South pole and, therefore, observes mostly the Northern sky \cite{icecube}. Furthermore, thanks to the excellent optical properties of the deep sea water, the angular resolution of ARCA is significantly better than that of IceCube, providing an unprecendeted sensitivity to point-like sources of neutrino and improved prospects to associate the neutrino sources with specific astronomical objects \cite{km3net_loi}.

The recent observation of a diffuse flux of cosmic neutrinos by IceCube \cite{icecube} promises an exciting future for ARCA. Indeed, the existence of PeV hadronic accelerators in the Universe has been proved, as well as the feasibility of observing neutrinos originating from such accelerators. However, many questions remain unanswered concerning the origin of the cosmic neutrinos, in particular, whether these neutrinos originate within or outside our Galaxy, what type of astrophysical objects and environments serve as the neutrino production sites, and how particles are accelerated to PeV energies. ARCA will allow to answer these questions by observing the neutrino sources and pinpointing their locations on the sky with an unprecedented accuracy. The ARCA research program also includes studies of the flavour composition of the cosmic neutrino flux, searches for transient neutrino sources, indirect dark matter searches, searches for exotic particles and Lorentz invariance violation. Further information on the ARCA physics case can be found in \cite{km3net_loi}.

The first two ARCA DUs were successfully deployed and connected to the shore station in December 2015 and May 2016, respectively. The recorded data demonstrate a satisfactory operation of all detector subsystems. A measurement of the atmospheric muon flux as a function of depth has been obtained showing a good agreement with expectations.

\section{Neutrino mass hierarchy measurement with ORCA}

Each of the three flavour eigenstates of neutrino ($\nu_e$, $\nu_\mu$, $\nu_\tau$) is a superposition of three mass eigenstates ($\nu_1$, $\nu_2$, $\nu_3$) with three different masses ($m_1$, $m_2$, $m_3$) \cite{pmns}. The neutrino mass ordering (“mass hierarchy”) is only partially constrained by existing experimental data. More specifically, only the relative ordering of the first and second neutrino mass eigenstates can be fixed ($m_1 < m_2$), based on observations of the Solar neutrino oscillations \cite{sno}. It is currently unknown whether the third neutrino mass eigenstate is heavier or lighter than the other two mass eigenstates. Given that the first squared mass splitting $\vert m_1^2 - m_2^2 \vert$ is much smaller than the the second one $\vert m_2^2 - m_3^2 \vert$, there are two possible orderings: $m_1 < m_2 < m_3$ (“normal mass hierarchy”) and $m_3 < m_1 < m_2$ (“inverted mass hierarchy”). The Standard Model has no particular preference for either type of the mass hierarchy, as it generally assumes neutrinos to be massless. However, the neutrino mass hierarchy may be related to the underlying structure of physics Beyond the Standard Model (BSM), and many BSM theoretical models provide specific predictions for the mass hierarchy \cite{nmh}. The mass hierarchy is also important for interpreting data from current experiments, as well as for planning of future experiments, for instance on double-beta decay. 

When neutrinos propagate in matter, such as inside the Sun or Earth, the flavour transition probabilities are modified due to coherent forward scattering of neutrino on atomic electrons -- the so-called Mikheyev-Smirnov-Wolfenstein (MSW) effect \cite{Wolfenstein,MikheevSmirnov}. Indeed, contrarily to the other flavours, the $\nu_e$ component can undergo charged-current (CC) elastic scattering interactions with the electrons in matter and, consequently, acquire an effective potential $A = \pm \sqrt{2} \, G_F N_e$, where $N_e$ is the electron number density of the medium, $G_F$ is the Fermi constant and the +(-) sign stands for $\nu_e$ ($\overline{\nu}_e$). For neutrinos propagating through the Earth, this effective potential leads to a resonant enhancement of $\nu_e$ appearance at energies E $\approx$ 3--8 GeV in the case of normal mass hierarchy. In the case of inverted mass hierarchy the resonance occurs instead in the $\overline{\nu}_e$ appearance channel. Hence, observations of the $\nu_e$ and/or $\overline{\nu}_e$ appearance at E $\sim$ 5 GeV allow to determine the neutrino mass hierarchy. 

\begin{figure}[t]
\centering
\includegraphics[height=6cm]{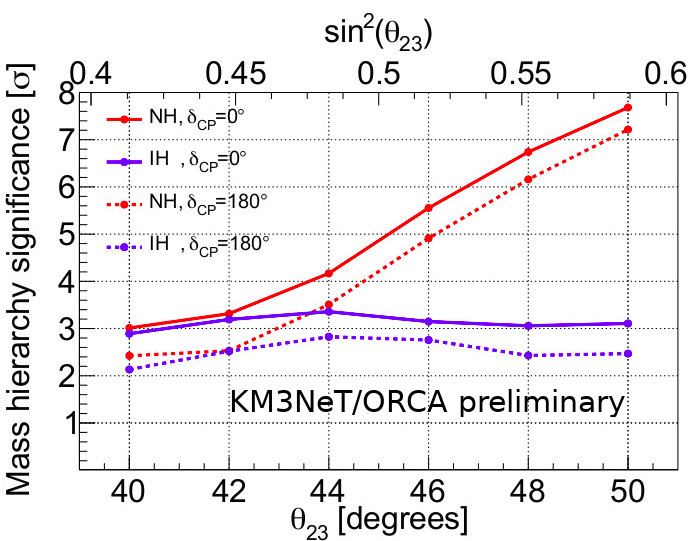}
\caption{The mass hierarchy sensitivity of ORCA after three years of operation, shown as a function of $\theta_{23}$ and for two values of $\delta_{CP}$.}
\label{orca_sensi}
\end{figure}

The primary mission of ORCA is to determine the neutrino mass hierarchy (normal vs. inverted). This will be accomplished via precision measurements of the atmospheric neutrino oscillations in the energy range where the MSW effect in the Earth (core, mantle and crust) has the strongest impact on the neutrino oscillation pattern (between 3 GeV and 20 GeV). Thanks to its large volume and low energy threshold ($\sim$3 GeV), ORCA will detect about 50000 atmospheric neutrinos every year. Most of these neutrinos will have muon or electron flavour. The relative abundance of muon and electron (anti-)neutrinos will be measured on a statistical basis using the characteristic features of muon neutrino charged current (CC) events, in particular the presence of a long muon track. Studies performed by the KM3NeT Collaboration suggest that at E$_{\nu}$ = 5~GeV the majority of $\nu_\mu$ CC events detected by ORCA can be correctly identified as muon neutrinos while $<$ 15\% of electron CC events are misidentified as muon neutrinos. The neutrino energy resolution of ORCA is about 30\% for both $\nu_\mu$ and $\nu_e$ CC events (at E$_{\nu}$ = 5 GeV). Based on latest sensitivity projections, a result with a 3 $\sigma$ significance on the mass hierarchy is expected after three years of data taking \cite{km3net_loi}. Under certain conditions (normal mass hierarchy, $\delta_{CP}$=0, $\theta_{23}$=50$^{\circ}$), the statistical significance may reach 6 $\sigma$ already after 3 years (see Fig.~\ref{orca_sensi}). ORCA will also provide improved measurements of the atmospheric neutrino oscillation parameters (e.g. $\theta_{23}$) and constraints on non-standard neutrino interactions, as well as sensitivity to astrophysical neutrino sources, dark matter, and other physics phenomena. The detector construction has recently started and is expected to be completed within about 4~yr, providing its first results on the mass hierarchy around 2023.

\section{Scientific case for a neutrino beam from Protvino to ORCA}

Big Bang cosmology requires that the CP symmetry is violated -- at least at some moment during the Big Bang -- in order to explain the dominance of normal matter over anti-matter in the present-day Universe \cite{Sakharov}. However, the origin of the CP violation remains yet unknown. Indeed, the only source of CP violation experimentally observed so far is a CP-violating phase $\delta$ in the mixing of quarks (J. Cronin \& V. Fitch, Nobel prize in physics, 1980). However, this CP phase alone is insufficient to explain the present-day abundance of matter in the Universe. Theoretically, the CP symmetry could also be violated by the strong interaction but this has not been experimentally observed, indicating that the effect is negligibly small or strictly zero. The only other known possible source of CP violation is the leptonic CP phase $\delta_{CP}$, associated with neutrino mixing [3]. (Additional CP phases may exist if neutrinos are Majorana particles -- these phases do not have observable consequences on neutrino oscillations and will be ignored in the following.)

The fact that all three neutrino mixing angles are different from zero leads to the possibility of probing $\delta_{CP}$ in the electron (anti-)neutrino appearance channel using the $\nu_\mu \rightarrow \nu_e$ ($\overline{\nu}_\mu \rightarrow \overline{\nu}_e$) transition. The most plausible setting for measuring this transition is offered by experiments with accelerator neutrino beams, which provide a clean source of muon neutrinos ($\nu_\mu$ or $\overline{\nu}_\mu$, depending on the chosen beam polarity). Such a measurement is actively pursued by several long-baseline neutrino experiments, but is so far limited by insufficient statistics. The only significant experimental constraints on $\delta_{CP}$ available so far come from the T2K and NOvA experiments \cite{t2k,nova,nova2018}. However in both cases the statistical significance of the result does not exceed 2 $\sigma$. Note that using atmospheric neutrinos ORCA has only a marginal sensitivity to $\delta_{CP}$.

It has recently been proposed that sending an artificial neutrino beam to ORCA would open up a new range of opportunities in neutrino oscillation research \cite{brunner2013}. In particular, such an experiment could provide a very sensitive probe of leptonic CP violation ($\delta_{CP}$). A neutrino beam of suitable energy could be produced at the Protvino accelerator facility, located 100 km South of Moscow, Russia. The core component of the accelerator facility is the U-70 synchrotron which accelerates protons up to 70 GeV, allowing for the production of an intense beam of neutrinos or anti-neutrinos with energies up to 7~GeV \cite{GarkushaNovoskoltsevSokolov2015}. With a 2590 km distance between Protvino and ORCA, the first neutrino oscillation maximum will be at 5 GeV, close to the matter resonance energy in the Earth crust (4 GeV) and within an energy range convenient for ORCA as well as for the accelerator. Such an experiment proposal was initially made in Ref. [12]. The proposal is now known as Protvino-to-ORCA experiment, or simply P2O.

The U-70 synchrotron currently operates at a time-averaged beam power of 15 kW. The accelerator power could be increased to $\approx$ 90 kW by means of a relatively inexpensive upgrade (new ion injection scheme and optimization of the accelerator cycle from 9 s down to $\approx$ 4.5 s). An upgrade up to 450 kW could be made possible by a new chain of booster accelerators \cite{omega}. A neutrino beam line will need to be constructed to produce a focused neutrino beam in the direction of ORCA. A near beam detector will also be needed, in order to accurately monitor the neutrino beam intensity, energy spectrum, and flavour composition before oscillations.

\begin{figure}
\centering
\includegraphics[height=7cm]{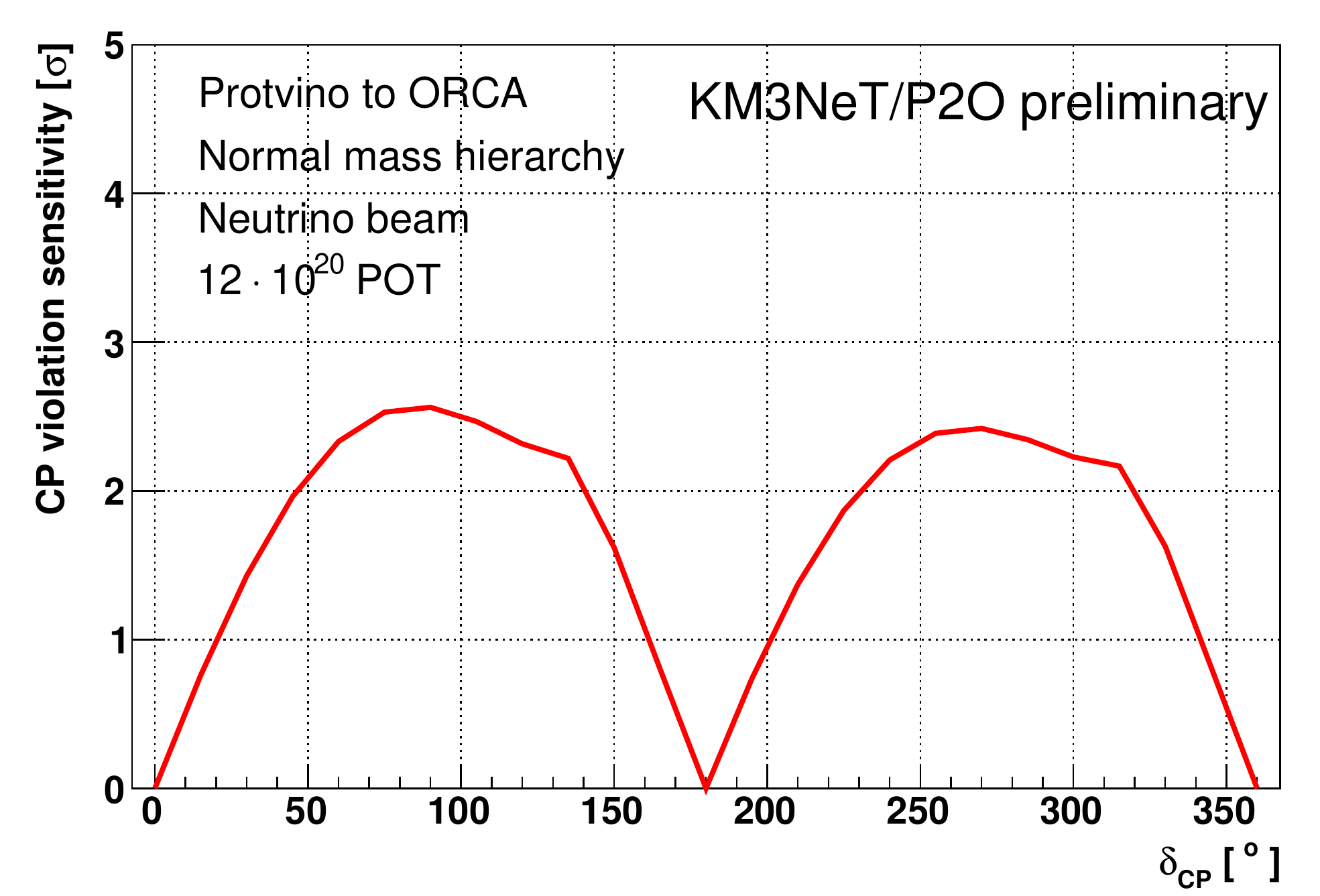}
\caption{Sensitivity of P2O to the presence of CP violation in the neutrino sector as a function of $\delta_{CP}$ after 3 yr of a 450 kW beam, or 15 yr of a 90 kW beam ($\delta_{CP}$ = 0 and 180$^{\circ}$ correspond to CP conservation). Normal mass hierarchy.}
\label{p2o_sensi_cpv}
\end{figure}

\begin{figure}
\centering
\includegraphics[height=7cm]{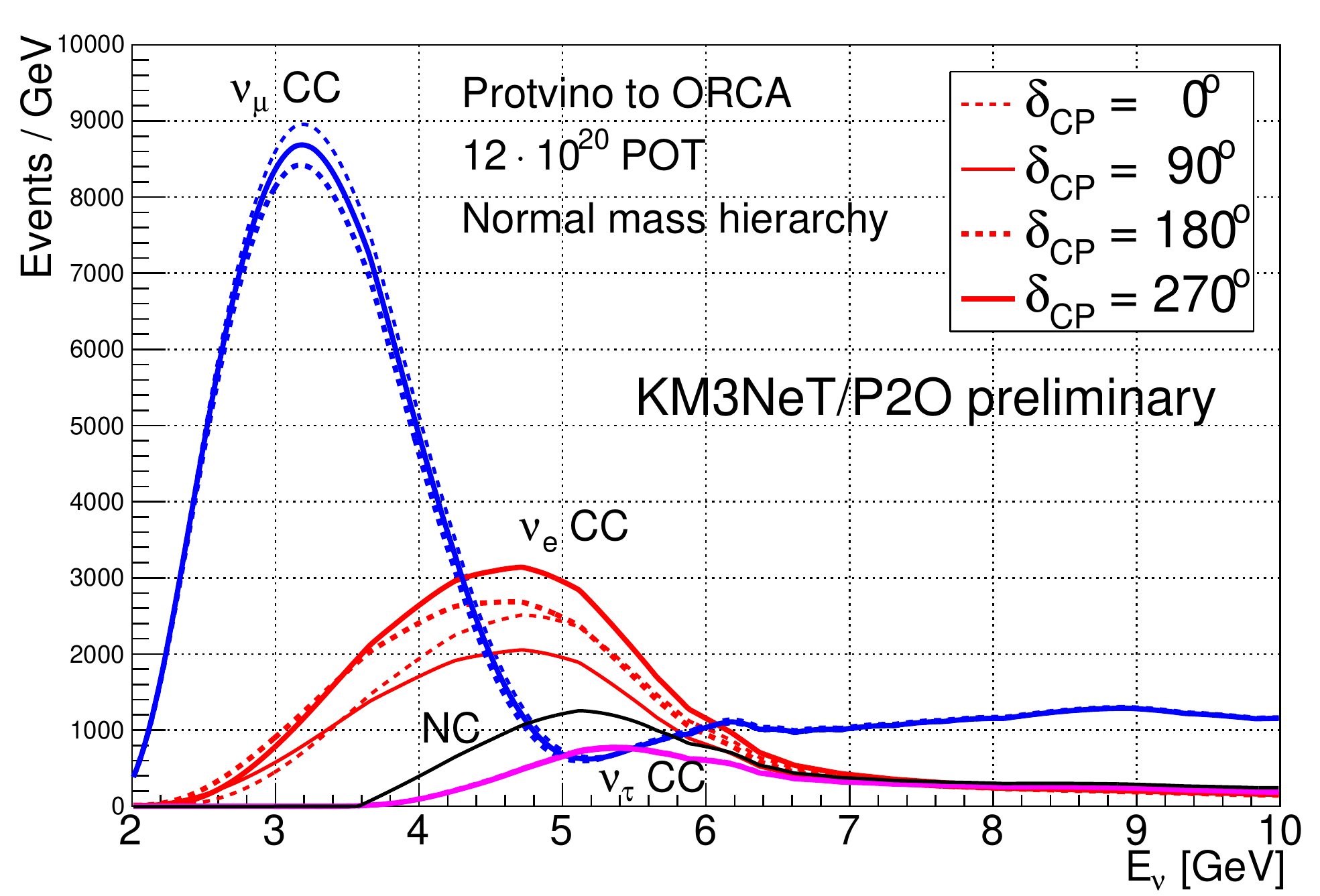}
\caption{
Number of neutrino events expected to be detected by ORCA after 3 years of running with the 450 kW beam, or 15 yr with the 90 kW beam. The case of normal neutrino mass hierarchy has been assumed. The x-axis shows the true neutrino energy.
The different line styles correspond to four different values of $\delta_{CP}$, as indicated in the legend.
The three groups of colored curves correspond to $\nu_\mu$ CC (blue), $\nu_e$ CC (red) and $\nu_\tau$ CC (magenta) interactions.
The black curve is for neutral current and contains the sum of three neutrino flavors (unaffected by oscillations).
}
\label{p2o_event_numbers}
\end{figure}

A preliminary study of the scientific potential of the P2O experiment suggests that the neutrino mass hierarchy could be determined with a 5--10 $\sigma$ significance after one year of running with a 450 kW beam (or five years with a 90 kW beam). This would provide a solid confirmation of the $\approx$ 3--5 $\sigma$ result expected to be obtained by ORCA using atmospheric neutrinos. It was also found that with 3 years (15 years) of the 450 kW (90 kW) beam the P2O experiment could provide a 2--3 $\sigma$ sensitivity to CP violation (see Fig.~\ref{p2o_sensi_cpv}), along with a better than 20$^{\circ}$ accuracy on the CP phase $\delta_{CP}$.  These estimates were obtained using a preliminary data analysis pipeline optimized for the atmospheric neutrino analysis. Potential analysis improvements can be added thanks to the known arrival direction and timing of the neutrino beam. Most of the sensitivity to the mass hierarchy and CP violation comes from the electron (anti-)neutrino appearance channel (see Fig.~\ref{p2o_event_numbers}). 

It is worth noting that a 90 kW beam will produce about 3000 neutrino events in ORCA every year. For comparison, the DUNE experiment, using a 1 MW beam in combination with a 40 kt detector over a 1300 km baseline, will detect only 1000 neutrino events per year. DUNE is expected to reach a 3 $\sigma$ sensitivity to CP violation using 15 yr of operation with the beam \cite{dune}. Hence, P2O would be competitive to DUNE, even when using a relatively low intensity beam (90 kW). P2O would also be complementary to DUNE and other long-baseline neutrino experiments, e.g. T2K/T2HK \cite{hyperk}, thanks to several unique characteristics of P2O: the highest neutrino event statistics, the longest baseline, and the highest energy of the oscillation maximum.

\section{Conclusion}
Two giant sea-water Cherenkov detectors, ORCA and ARCA, are now under construction by the KM3NeT Collaboration at two sites in the Mediterranean Sea. ORCA will allow for the study of the atmospheric neutrino oscillations and determine the neutrino mass hierarchy, while ARCA will open a new chapter in high energy neutrino astronomy. Sending an accelerator neutrino beam from Protvino (Russia) to ORCA promises exciting prospects for studies of leptonic CP violation ($\delta_{CP}$), while also providing a robust cross-check on the mass hierarchy. Such an experiment, dubbed P2O, would be complementary to and competitive with other present and future long-baseline experiments, in particular DUNE and T2K/T2HK. The P2O project will require an upgrade of the U-70 accelerator complex at Protvino to serve as a high-intensity neutrino source. Additionally, a near detector would be needed in the vicinity of the accelerator complex for accurate monitoring of the intensity and flavour composition of the neutrino beam.


\end{document}